\documentstyle[nato,numreferences]{crckapb}

\def\Journal#1#2#3#4{{#1} {\bf #2}, #3 (#4)}

\def\NPB{{\em Nucl. Phys.} B}
\def\PLB{{\em Phys. Lett.}  B}
\def\PRL{\em Phys. Rev. Lett.}
\def\PRD{{\em Phys. Rev.} D}
\def\PRP{{\em Phys. Rept.}}

\def\CMP{\em Commun. Math. Phys.}
\def\JPSJ{\em J. Phys. Soc. Jpn.}
\def\PRSL{\em Proc. R. Soc. London, Ser. A}
\def\NPROC{\em Nucl. Phys. Proc. Suppl.} 

\def\be{\begin{equation}}
\def\ee{\end{equation}}
\def\bea{\begin{eqnarray}}
\def\eea{\end{eqnarray}}
\def\sla{\raise.15ex\hbox{$/$}\kern-.57em}
\def\Tr{{{\rm\bf Tr}\ }}
\def\tr{{{\rm\bf tr}\ }}
\def\Ker{{{\rm\bf Ker}\ }}
\def\span{{{\rm\bf span}\ }}
\def\a{{{\rm ad} }}
\def\E{{{\rm \bf E} }}

\begin{opening}
\title{Chiral Symmetry Outside Perturbation Theory${}^1$}

\author{H. Neuberger}
\institute{Department of Physics and Astronomy\\
Rutgers University\\
Piscataway, NJ 08855\\
U.S.A.}

\end{opening}

\runningtitle{Chiral Symmetry Outside Perturbation Theory}

\begin{document}

\begin{abstract}
Within the overlap framework, I derive the main formulae
one finds today in papers touting a ``new approach'' 
to the regularization of chiral gauge theories. 
My main objective is to clear up an unhealthy confusion about
how many successful approaches to regulate chiral fermions on 
the lattice there really are: At the moment, there is only one,
the overlap, and finding a genuinely different approach 
is an important and completely open problem. 
\end{abstract}

\vspace{-11.0cm}
\begin{flushright}{\normalsize RUHN-99-6}\\
\end{flushright}
\vspace*{9.5 cm}

\section{Introduction}
\addtocounter{footnote}{1}
\footnotetext{Contribution to the proceedings of the workshop 
``Lattice fermions and the structure of the vacuum'', 5-9 October, 1999,
Dubna, Russia.}
The talk I delivered at this workshop had substantial overlap with 
talks I gave at Lattice'99 \cite{Npisa} and at Chiral'99 \cite{Nchiral}. 
To avoid repetition, this write-up is restricted to technical points which
were neither covered orally, nor in the above mentioned written contributions.

I shall show explicitly how the main formulae one finds today in papers 
touting a ``new approach'' to the regularization of chiral gauge theories 
\cite{Lpisa,Nboulder,Label,Lnonabel} are directly and straightforwardly 
derived from the overlap. Thus, there really is no ``new'' approach and
there are fewer new results than a superficial reading of the above
papers would indicate. Whether one likes to start from the Ginsparg-Wilson 
\cite{gw} relation or from the overlap one arrives at the same algebraic
setup. The GW relation, by itself, does not guarantee the
right dynamics, and one needs to fulfill extra side conditions in order
to really get chiral fermions \cite{Nchiral}. The GW relation 
became fashionable in January 1998.
The overlap Dirac operator first appeared 
in \cite{plbovlapdirac} (posted July, 1997) 
and its connection to the GW relation was pointed 
out in \cite{prdvector} (posted October, 1997). 
This write-up consists of a collection
of formulae related to overlap chiral fermions on the lattice
with their derivations; each derivation amounts to a little exercise 
in linear algebra and is quite trivial. 

My main objective in putting this summary in print has been stated in the
abstract. To be sure, let me add that {\it some} 
new results have indeed been obtained recently: 
During the last two years a certain amount of progress has taken place 
on the mathematical 
question of fine tuning the phase of the overlap to eliminate small gauge 
breaking effects when anomalies cancel.

\section{Notation}
Our focus is on lattice Dirac fermions defined on a four dimensional
hypercubic lattice in a background of $SU(n)$ lattice gauge fields. 
These fermions live in a finite complex vector space of even dimension $N$.
Elements in this space will be denoted as ${\vec v}$. Components of these
vectors will labeled by combined indices, $I,~J$, with the convention
$I=(x,i,\alpha), ~J=(y,j,\beta)$, where $x, ~y$ label sites, $i,~j$
group indices and $\alpha, ~\beta$ spinor indices. Operators will
be represented by matrices with matrix indices appearing as subscripts.
If only site indices appear the group and spinor indices are to be
understood as suppressed. Trace operations either operate on all indices
($\Tr$) or only on a restricted set, typically excluding sites ($\tr$).
In addition to square matrices we shall often employ rectangular ones,
dimensioned as (number of rows) $\times$ (number of columns).

Two main reflection operators act on the vectors ${\vec v}$:
$\epsilon$ and $\epsilon^\prime$. A reflection is a unitary-hermitian
operator. Equivalently, one can think about the associated projectors 
$P={1\over 2}(1-\epsilon)$, $P^\prime ={1\over 2} (1-\epsilon^\prime )$ 
as fundamental. The Kato \cite{kato} pair $h={1\over 2} (\epsilon+
\epsilon^\prime )$ and $s={1\over 2} (\epsilon-\epsilon^\prime )$ is 
algebraically characterized \cite{avrona} by $h^2 + s^2 =1$ 
and $\{h,s\}=0$, 
where the anticommutator can be viewed as a version of the Ginsparg-Wilson 
relation, or a lattice version of chiral symmetry \cite{Nchiral}. 
A central role is 
played by the overlap Dirac operator 
\cite{plbovlapdirac} 
$D_o ={1\over 2} (1+\epsilon^\prime 
\epsilon)=\epsilon^\prime h $. Nothing of principle 
is lost by setting $\epsilon^\prime = \gamma_5$. Some trivial 
identities are listed below:
\begin{equation}
\epsilon^\prime h = h \epsilon ,~
P^\prime h = h P,~\epsilon h = h \epsilon^\prime ,~
P h = h P^\prime.
\end{equation}
In the above one can replace $h$ by $h^{-1}$ when the inverse exists.
Similar equations, up to signs, are obeyed by $s$. 
Another trivial identity is $h=1-P-P^\prime$. From it one
derives $P^\prime h^{-1} P^\prime =-P^\prime$ and similarly  
\begin{equation}
Ph^{-1}P=-P 
,~(1-P^\prime ) h^{-1}(1-P^\prime ) = 1- P^\prime,
~(1-P) h^{-1} (1-P) = 1-P.
\end{equation}
The inverse of the overlap Dirac operator obeys:
\begin{equation}
D_o^{-1} -1 =
{{1-\epsilon^\prime \epsilon}\over{1+\epsilon^\prime \epsilon}}
=sh^{-1}=-h^{-1}s .
\end{equation}
The last expressions obviously anti-commute with $\epsilon^\prime$ and are
anti-hermitian. 

A second quantized notation will also be employed when appropriate:
We imagine dealing with a system of $N$ noninteracting 
fermions represented by a $2^N$ dimensional Fock space. The elements
making up the Fock space are
superpositions of anti-symmetrized direct 
products of single particle states obtained
using any basis of the original complex N-dimensional space one chooses.
One assumes a standard basis relatively to which standard fermionic
creation/annihilation operators $a_I^\dagger$/$a_I$ are defined. 
To distinguish vectors in the Fock space from vectors in other spaces we 
shall use Dirac bra-ket notation for second quantized states only.

In the overlap one needs to fill all the negative energy states of
$\epsilon$. They generate 
$\Ker(1+\epsilon) = \span\{{\vec v}_i |i=1,2,...,N_v\}$.
$\epsilon$ depends on the gauge fields $U_\mu (x)$ and so do the 
orthonormal vectors ${\vec v}_i$. Similarly one introduces
$\Ker(1-\epsilon) = \span\{{\vec w}_i |i=1,2,...,N_w\}$, with
$N_v + N_w =N$. Appending a prime, similar objects are introduced for
$\epsilon^\prime$, but now there is no gauge field dependence and
$N_{v^\prime}=N_{w^\prime}={1\over 2}N$. 
The Dirac sea state corresponding to occupying all ${\vec v}_i$
single fermion states will be denoted by $|v \rangle$.

It is convenient to collect all the vectors ${\vec v}_i$ into
an $N\times N_v$ matrix $v=( {\vec v}_1 ,{\vec v}_2,....{\vec v}_{N_v} )$
and do the same for similar collections of vectors. Then:
\begin{eqnarray}
&P=vv^\dagger,~~v^\dagger v = 1,~~1-P=ww^\dagger,~~w^\dagger w=1 ,\\&
~Pv=v,~P^\prime v^\prime = v^\prime,
~Pv^\prime = D_o^\dagger v^\prime, ~P^\prime v = D_o v .
\end{eqnarray}

Starting from identities like $v^{\prime\dagger} h^{-1}
v^\prime = v^{\prime\dagger} 
P^\prime h^{-1} P^\prime v^\prime$ we get:
\begin{equation}
v^{\prime\dagger} h^{-1} 
v^\prime = v^\dagger h^{-1} v=-1 ,
~w^{\prime\dagger} h^{-1} w^\prime = w^\dagger h^{-1} w =1. 
\end{equation}

\section{Overlap determinant and propagator}
\subsection{Chiral case}

Pauli's statistics for fermions implies 
\begin{equation}
\langle v^\prime |v \rangle = \det M_R ,~~~M_R=v^{\prime \dagger}v .
\end{equation}
This is the overlap and gives the lattice chiral fermion determinant.
By convention, it is associated with right handed Weyl fermions ($R$). 
The inverse of $M_R$ is 
\begin{equation}
M_R^{-1}=-v^\dagger h^{-1} v^\prime 
\end{equation}
because $-v^{\prime\dagger}vv^\dagger h^{-1} v^\prime =
-v^{\prime\dagger} P h^{-1}v^\prime = -v^{\prime\dagger} h^{-1}v^\prime =1$. 

The matrix $M_R$ can be rewritten in an artificial way to look dependent
also explicitly on the overlap Dirac operator:
\begin{equation}
M_R=v^{\prime\dagger} D_o v .
\end{equation}
The equation holds by taking the factors $\epsilon^\prime$ and $\epsilon$
in $D_o$ to act left and right respectively. 
This is equation (18) in \cite{Lpisa}.
It is strange that the simpler overlap form is never even mentioned in
\cite{Lpisa,Nboulder,Label,Lnonabel}.

$M_R^{-1}$ is not directly equal to the R-fermion propagator, $G^R$. 
Unlike $M_R$, $G^R$ is an operator on the original vector
space. Since it represents the propagation of $R$-Weyl fermions
but acts in a space that accommodates Dirac fermions it is appropriately
rank deficient. In the overlap construction 
(equation (5.19) in \cite{npblong}) the propagator is found to be
\begin{equation}
G_{JI}^R ={{\langle v^\prime | a_I^\dagger a_J | v\rangle }\over
{\langle v^\prime | v \rangle }},~~~G^R =vM_R^{-1}v^{\prime\dagger}.
\end{equation}
When it exists, $G^R$ has rank ${N\over 2}$ and is given
by $G^R = -Ph^{-1}P^\prime =-h^{-1} P^\prime$. 
Since $\epsilon^\prime P^\prime = - P^\prime$
\begin{equation}
G^R =P h^{-1}\epsilon^\prime P^\prime = PD_o^{-1} P^\prime .
\end{equation}
The last expression is equation (17) in \cite{Lpisa}, 
but $G^R = -h^{-1} P^\prime $ is simpler.

The formula $G^R =-h^{-1}P^\prime$ is important because it 
makes it explicit that $G^R$ transforms covariantly under 
gauge transformations. 
This covariance is self-evident in the second quantized expression
in terms of operators. 
Thus, $\tr \Gamma G_{x,y}^R W_{y,x}$, 
where $W_{y,x}$ is a Wilson line operator
connecting sites $x$ and $y$ and $\Gamma$ acts only on spinorial indices, 
is gauge invariant although $\det M_R$ is generically not. In this aspect the
overlap is different from earlier attempts to put chiral fermions
on the lattice, break gauge invariance and restore it subsequently
by gauge averaging. In the overlap gauge averaging cannot destroy
the perturbative masslessness of the fermions, so, for instance,
the counter example of Testa \cite{testakorea} does not apply. 

For $L$-fermions we introduce $M_L = w^{\prime\dagger}w$ with
inverse $M_L^{-1} = w^\dagger h^{-1} w^\prime$ and
propagator $G^L = h^{-1} (1-P^\prime )=
(1-P)h^{-1}$, obtained
by reversing the signs of $\epsilon$ and $\epsilon^\prime$ \cite{npblong}.
This leads to
\begin{equation}
G^R + G^{L\dagger}=(1-P-P^\prime ) h^{-1} =1 .
\end{equation}
To better match continuum properties \cite{npblong} we define the overlap
chiral external propagators $G_o^R =G^R-{1\over 2}$ and 
$G_o^L =G^L-{1\over 2}$. This corresponds to replacing
$a_I^\dagger a_J$ by ${1\over 2} (a_I^\dagger a_J -
a_J a_I^\dagger )$ in the second quantized formula (equation (5.22)
in \cite{npblong}). One can take an $R$-$L$ combination to propagate
with $G_o^R + G_o^L \equiv G^V$. 
\begin{equation}
G^V = -h^{-1} P^\prime +h^{-1}(1-P^\prime ) -1 = h^{-1}\epsilon^\prime -1
=D_o^{-1}-1=sh^{-1}
\end{equation}
anti-commutes with $\epsilon^\prime$; however, $G^R + G^L = D_o^{-1}$
does not.

\subsection{Vector-like case}
The fermion determinant in the vector-like case is $|\det M_R |^2 =
\det v^{\prime\dagger} P v^\prime$. Hence $|\det M_R |^2 =$
\begin{equation}
\det [1-v^{\prime\dagger} (1-P) v^\prime ] =
e^{\sum_{m=1}^\infty {1 \over m} \Tr [P^\prime (1-P)]^m }=
\det [1-P^\prime (1-P)].
\end{equation}
Similarly, one shows $|\det M_R |^2 = |\det M_L |^2 = $
\begin{equation}
\det [1-(1-P)P^\prime ]=\det [1-P(1-P^\prime)]=
\det [1-(1-P^\prime ) P] .
\end{equation}
Comparing the matrix elements 
between the bases $\{ {\vec v}_i^\prime , 
{\vec w}_j^\prime \}$ and  $\{ {\vec v}_i , {\vec w}_j \}$
of the above combinations of projectors
with those of $D_o$ we obtain
\begin{equation}
|\det M_R |^2= |\det M_L |^2=\det D_o .
\end{equation}
Since $\det \epsilon^\prime =1$,  $\det D_o=\det h$.
The propagator on internal fermion lines is given by $D_o^{-1}=G^R
+G^L$. Only on external fermion lines can one use $G^V =
D_o^{-1} -1$ and preserve na{\" \i}ve chiral symmetry exactly
\cite{Neubboulder}.

\section{Consistent, Covariant Currents, Anomalies and Topology}

To get currents one computes the first order
variation of the chiral determinant with respect to
the gauge fields. The precise form of the variation 
($\delta$) is not important here. The variation can be naturally
(geometrically) decomposed into two terms \cite{npbfirst,seifcons,Ngeom}.
\begin{equation}
\delta \log \langle v^\prime | v \rangle = 
{{\langle v^\prime | \delta v_\perp \rangle} \over
{\langle v^\prime | v \rangle}}+\langle v | \delta v\rangle ,
\end{equation}
to isolate the dependence on the phase choice for $|v\rangle$ into
the local last term. The main point is that because of the phase
ambiguity the component of $|\delta v\rangle$ along $|v\rangle$ is
not determined by $\delta\epsilon$ but $|\delta v_\perp \rangle$ is.
The first term, being phase choice independent
can be made to transform covariantly under gauge transformations with
an appropriate choice of the variation $\delta$. This term is
nonlocal in gauge fields and defines the covariant current. The
sum of both terms is also a current and, assuming a single valued (as
a function of the gauge background) choice
of the second quantized states $|v\rangle$ has been made, gives the 
consistent current with an appropriate choice of $\delta$. 
The difference,
denoted by $\Delta J$ in the continuum \cite{barzum}, is the last
term. This last terms is recognized as the Berry connection 
\cite{berry}: Under
a change of phase $|v\rangle \rightarrow e^{i\Phi (U)} |v\rangle$
it changes additively by $i\delta \Phi$, but it contains invariant
information in associating a Berry phase $\Phi (C)$
with every closed contour $C$ in gauge field space. 

Let us now translate back to first quantized language:
\begin{equation}
\langle v | \delta v \rangle =
\sum_{i=1}^{N_v} {\vec v}_i^\dagger \delta
{\vec v}_i = \Tr v^\dagger \delta v.
\end{equation}
For simplicity, we assume $\langle v^\prime |
v\rangle \ne 0$ (which also implies $N_v = {N\over 2}$). 

The variation of $\log \langle v^\prime |
v\rangle $ is just $\Tr M^{-1}\delta M =
- \Tr v^\dagger h^{-1} v^\prime v^{\prime\dagger}
\delta v = -\Tr v^\dagger h^{-1} \delta v =-\Tr v^\dagger h^{-1} \delta (Pv)
$ which is equal to:
\begin{equation} 
-\Tr h^{-1} P^\prime \delta P -\Tr v^\dagger P h^{-1} P \delta v=
\Tr v^\dagger \delta v + \Tr P^\prime \delta h h^{-1}
\end{equation}
The last term can be rewritten as
\begin{equation}
\Tr P^\prime \delta h h^{-1}=-\Tr \epsilon^\prime \delta h h^{-1} P^\prime
=\Tr \delta D_o G^R ,
\end{equation}
leading to an expression for the covariant current:
\begin{equation}
{{\langle v^\prime | \delta v_\perp \rangle} \over
{\langle v^\prime | v \rangle}} = 
\Tr P^\prime \delta h h^{-1} =
\Tr P^\prime \delta D_o D_o^{-1} = 
\Tr \delta D_o G^R .
\end{equation}
One should not forget however that $G^R $ is not $D_o^{-1}$. 
The above equation contains formula (21) in \cite{Lpisa}. 
In second quantized notation we have:
\begin{equation}
{{\langle v^\prime | \delta v_\perp \rangle} \over
{\langle v^\prime | v \rangle}} = 
{{\langle v^\prime |a_I^\dagger (\delta D_o )_{IJ} a_J | v \rangle} \over
{\langle v^\prime | v \rangle}}  .
\end{equation} 
So, all we calculated
is the bilinear numerical kernel giving the current operator
associated with varying the fermion induced effective action 
in second quantized language. Actually, the second quantized form
is advantageous in topologically nontrivial backgrounds. There,
to make the expression meaningful one needs to insert some operator
of the 't Hooft vertex type in the numerator (the denominator
is ``canceled'' by the fermion determinant factor). Thus one
can consider correlators between 't Hooft vertices and covariant
currents. 

To get the covariant anomaly we choose the variation to be an
infinitesimal gauge transformation with parameters $\omega^a (x)$
at site $x$ where $a$ labels the $n^2-1$ hermitian generators
$t^a_r$ acting on the fermions which are in a representation (possible
reducible) $r$. The site diagonal matrix ${\cal G}_{I,J} = 
i \omega^a (x) \delta_{xy} (t^a_r)_{ij} \delta_{\alpha\beta}$ 
represents the transformation
in fermion space. Thus, $\delta h = [{\cal G}, h ]$,
reflecting the covariance of $h$.

Starting from $\Tr [{\cal G}, h] h^{-1} P^\prime = 
$
\begin{equation}
\Tr  {\cal G} P^\prime - \Tr h {\cal G} h^{-1} P^\prime =
{1\over 2} \Tr {\cal G} - \Tr h {\cal G} P h^{-1}
={1\over 2} \Tr {\cal G} \epsilon 
\end{equation}
we obtain for the covariant anomaly:
\begin{equation}
{i\over 2} \sum_x \omega^a (x) \tr (t^a_r \epsilon_{x,x} )\equiv
i\sum_x \omega^a (x) \triangle^a (x) .
\end{equation}
In the equations above 
the tracelessness of $\epsilon^\prime$ in spinor space was used. The
anomaly $\triangle^a (x)$ can also be trivially rewritten as
\begin{equation}
\triangle^a (x) =  \tr \epsilon^\prime t^a_r (D_o)_{x,x} .
\end{equation}
This is equivalent to equation (24) in \cite{Lpisa} (there is
a factor of two difference stemming from the choice in \cite{Lpisa}
to write the GW relation as $\{ D,\gamma_5 \} =D \gamma_5 D$ rather
than the overlap form $\{ D_o ,\gamma_5 \} =2 D_o \gamma_5 D_o $).
The anomaly equation is meaningful even when $D_o$ is not invertible.

The topological charge $Q$ in the overlap has been long known 
\cite{npbfirst,prl,npblong} to be given
by ${1\over 2} \Tr \epsilon$. Since
\begin{equation}
Q={1\over 2} \Tr \epsilon ={1\over 2} \sum_x \tr \epsilon_{x,x} = \sum_x 
\triangle^{U(1)} (x) ,
\end{equation}
we see the expected relation between the anomaly and the index. 

\section{Berry phase issues}
\subsection{Berry's curvature}
We already introduced Berry's connection ${\cal A}_i=
\langle v | \delta_i v \rangle = \Tr v^\dagger 
\delta_i v$. Under a phase change 
$|v\rangle \rightarrow e^{i\Phi} |v\rangle $
${\cal A}_i \rightarrow {\cal A}_i +i\delta_i \Phi$, but
the associated (abelian) Berry curvature ${\cal F}_{12}$ 
is unaffected \cite{Ngeom}:
\begin{equation}
{\cal F}_{12}=\delta_1 {\cal A}_2 - ({1\leftrightarrow 2})=
\Tr \delta_1 v^\dagger \delta_2 v -({1\leftrightarrow 2}).
\end{equation}
The phase freedom of the second quantized state $|v \rangle$
amounts to an arbitrary {\it unitary} ${\cal O}$ 
rotation among the first quantized states making up the Dirac sea. 
$\Tr v^\dagger \delta v \rightarrow 
$
\begin{equation}
\Tr {\cal O}^\dagger v^\dagger \delta (v {\cal O}) 
=\Tr v^\dagger \delta v + \Tr {\cal O}^\dagger \delta {\cal O} =
\Tr v^\dagger \delta v + \delta \log \det {\cal O}.
\end{equation}
Hence $e^{i\Phi}=\det {\cal O}$, and the rest of ${\cal O}$ is irrelevant.
The intrinsic meaning of Berry's curvature is made explicit
by expressing it in terms of the projectors only. 
The relevant formula is well known \cite{avronprl} and has
been used in \cite{Ngeom}. Start from
\begin{equation}
\Tr P \delta_1 P \delta_2 P = \Tr vv^\dagger (\delta_1 v v^\dagger
+v \delta_1 v^\dagger ) (\delta_2 v v^\dagger
+v \delta_2 v^\dagger )
\end{equation}
and expand the right hand side into a sum of four terms. Three of them are
symmetric in the $1,2$ indices and the one which is not is
$\Tr \delta_1 v^\dagger \delta_2 v$. Observing that $\Tr \delta_1 P
\delta_2 P$ is also symmetric we can replace every 
$P$ by a term $-{1\over 2}\epsilon$, obtaining
\begin{equation}
{\cal F}_{12} = -{1\over 8} 
\Tr \epsilon [\delta_1 \epsilon , \delta_2 \epsilon].
\end{equation}
This formula, with projectors in the reflections' stead, 
is eq (3.21) in \cite{Label}.

Let us now choose explicit formulae for the variations. 
We parameterize the group by real coordinates
$\xi^a$, so there is one set of $\xi's$ for every link: 
$\delta X = {{\partial X}\over {\partial \xi^{a} (x,\mu ) }} 
\delta \xi^a (x, \mu)$, 
with the summation convention acting on $a,x,\mu$. 
One would rather use vector fields
with nicer transformation properties than those of 
${{\partial }\over {\partial \xi^a (x,\mu ) }}$. Focusing
for the moment on a single copy of the group 
we opt to use the globally defined left invariant
vector fields $I_b = u^a_b (\xi) {{\partial }\over {\partial \xi^a}}$.
The $I_a (\xi )$ vector fields represent the Lie Algebra 
(with real structure constants $f_{ab}^c$) acting on the group
manifold: $[I_a (\xi ), I_b (\xi ) ] = f_{ab}^c I_c (\xi )$.
The main point about the introduction of the real 
matrix $u(\xi )$ \cite{oraifeartaigh} is that, for an arbitrary
group element parameterized by $\xi$, $g(\xi)$, we have: 
\begin{equation}
u^a_b (\xi) {{\partial g (\xi )}\over {\partial \xi^a }}=
g(\xi ) \left ( {{\partial g (\xi^\prime )}\over {\partial \xi^{\prime b}}}
\right )_{\xi^\prime =0}.
\end{equation}
This proves that for any two fixed group elements $g_1$ and $g_2$
\begin{equation}
I_a (\xi ) ( g_1 g(\xi ) g_2 ) = ( g_1 g(\xi ) g_2 )
(g_2^{-1} \left ({{\partial g (\xi^\prime )}\over {\partial \xi^{\prime a}}}
\right )_{\xi^\prime =0} g_2 )
\end{equation}
showing that gauge transformations act linearly on the covariant
currents defined below.

Let $\E_a (x,\mu )  = I_a ( \xi (x,\mu ) )$ with
$[\E_a (x,\mu ) , \E_b (y,\nu )]=f_{ab}^c \delta_{xy}
\delta_{\mu\nu} \E_c (x,\mu )$.
\begin{equation}
{\cal A}_a (x,\mu )=\langle v | \E_a (x,\mu ) v\rangle = 
\Tr v^\dagger \E_a (x,\mu )v 
\end{equation}
\begin{equation}
J_a^{\rm cov} (x,\mu ) = 
{{\langle v^\prime | [(\E_a (x,\mu ) v ]_\perp \rangle}
\over
{\langle v^\prime | v \rangle}} =
\Tr [\E_a (x,\mu ) D_o ] G^R .
\end{equation}
The antisymmetric tensor over field space $
{\cal F}_{ab} (x,\mu ;y,\nu ) $ is given by 
\begin{equation}
 \langle \E_a (x,\mu ) v | 
\E_b (y,\nu )v\rangle =
\Tr [ \E_a (x,\mu ) v^\dagger] [\E_b (y,\nu ) v ], ~{\rm antisymmetrized.}
\end{equation}
\begin{eqnarray*}
{\cal F}_{ab} (x,\mu ;y,\nu ) =
\Tr P [\E_a (x,\mu )P ,
\E_b (y,\nu ) P ]  =
\E_a (x,\mu ){\cal A}_b (y,\nu ) 
\end{eqnarray*}
\begin{equation}
- \E_b (y,\nu ){\cal A}_a (x,\mu )
-f_{ab}^c \delta_{xy} \delta_{\mu\nu} {\cal A}_c (x,\mu ).
\end{equation}
My notation and definitions \cite{Ngeom} 
are more general than in \cite{Lnonabel}, so that they 
apply to any Lie Group and any representation. In the $SU(n)$ case
with matrices in the fundamental the above reduces 
to equation (8.1) in \cite{Lnonabel}. 

\subsection{$Z(2)$ anomaly}

The gauge group $SU(2)$ with fermions in non-integral representations
is an interesting case because one can take the basic vector space
over the reals \cite{Nrealsu}. This makes Berry's connection and curvature
vanish; still the state $|v\rangle$ has a sign ambiguity - the $U(1)$
bundle of the complex case has been replaced by a $Z(2)$ bundle. If 
the $Z(2)$ bundle is twisted (like a M{\" o}bius strip), one cannot
find a single valued smooth definition for the states $|v\rangle$. If
one could, the signs of all $|v\rangle$'s would be determined by smoothness
up to an irrelevant overall sign. Then, lattice gauge transformations, which
can always be smoothly deformed to the identity, could not
induce sign changes and the chiral determinant would be gauge invariant.

In the continuum Witten \cite{witten} 
has shown that a gauge invariant formulation
for representations ${1\over 2},{5\over 2},...$ is impossible because
the space of {\it gauge orbits} (unlike the space of gauge fields) is
multiply connected: There are two classes of closed curves and 
the chiral determinant when taken round a curve in the nontrivial class
can change sign, so is not single valued, making a gauge invariant
formulation impossible. A nontrivial curve is obtained from an open
path connecting a gauge field configurations to its gauge transform 
by a gauge transformation which cannot be smoothly deformed to unity, 
and subsequently going to orbit space. 

Although Witten's anomaly is reproduced on the lattice \cite{Nztwoano},
the realization cannot be by exactly the same mechanism 
as in the continuum because on 
the lattice the space of all gauge transformations over the torus is
connected (any lattice gauge transformation can be smoothly deformed to
unity), while in the continuum it is not and this is the heart
of the matter \cite{witten}. The space of gauge orbits is multiply
connected also on the lattice, but this time as a consequence of the
space of gauge fields itself being already multiply connected (unlike
in the continuum) as a result of a necessary gauge 
invariant excision of backgrounds where
$\epsilon$ cannot be unambiguously defined \cite{Nztwoano}.

\subsection{Adiabatics}

It is well known that Berry's phase \cite{berry} is captured by adiabatic
Hamiltonian evolution. The quantum mechanical adiabatic theorem
states that in the adiabatic limit Hamiltonian evolution can be
replaced by a geometric evolution with an operator introduced
by Kato \cite{katoad} over forty years ago. The geometric evolution
amounts to parallel transport with Berry's connection \cite{simon}.
Considering a path in gauge field space 
parameterized by $t_1 \leq t\leq t_2$ this parallel transport means
\begin{equation}
\langle v_\a | {\dot v}_\a \rangle =0 ,
\end{equation}
where we assumed $\langle v_\a | v_\a \rangle\equiv 1$ and the dot denotes 
a derivative with respect to $t$. 

If we set $|v_\a (t_1 ) \rangle = |v(t_1 )\rangle$, where the 
``time'' argument identifies a gauge field configuration, 
adiabatic evolution means
that at all $t$ $|v_\a (t ) \rangle$ will be equal to 
$|v (t ) \rangle$ up to phase, the phase of
$|v_\a (t ) \rangle$ being fixed relatively to that of 
$|v (t ) \rangle$ by the above law of parallel transport. 

In first quantized language the phase arbitrariness of the states
$|v \rangle$ becomes the arbitrariness of $v$ under a unitary rotation
by ${\cal O}$. If all states ${\vec v}_i,~{\vec w}_j$ are evolved
adiabatically along our path they become 
${\vec v}_{\a, i},~{\vec w}_{\a, j} $ with the associated matrices related by
\begin{equation}
v(t) = v_\a (t ) {\cal O}_v (t),~~~~w(t) = w_\a (t ) {\cal O}_w (t),
\end{equation}
with initial conditions ${\cal O}_v (t_1 )=1$ and 
${\cal O}_w (t_1 )=1$. The ${\cal O}$ matrices are uniquely
defined as a function of $t$ by the law of adiabatic transport:
\begin{equation}
v_\a^\dagger (t) {\dot v}_\a (t)=0 ,~~~~w_\a^\dagger (t) {\dot w}_\a (t) =0 .
\end{equation}
The unitary transformation $K(t)$ producing the time evolution of 
the adiabatically evolved basis is given by 
\begin{equation}
v_\a (t) v^\dagger (t_1 ) + w_\a (t) w^\dagger (t_1 ) =
v(t ) {\cal O}_v^\dagger (t ) v^\dagger (t_1 ) +
w(t ) {\cal O}_w^\dagger (t ) w^\dagger (t_1 ) .
\end{equation}
Kato defined $K(t)$ by the following first order
differential equation
\begin{equation}
{\dot K} = [{\dot P }, P ] K,~~~~K(t_1 ) =1 .
\end{equation}
The formula is proven starting 
from $P(t) =v_\a (t) v_\a^\dagger (t)$ and observing: 
\begin{equation}
{\dot P} P v_\a = {\dot v}_\a ,~~ 
P{\dot P} v_\a =0,~~ P{\dot P} w_\a = -{\dot w}_\a . 
\end{equation}
$| v_\a (t ) \rangle = e^{i\Phi (t) } |v (t ) \rangle$
and, as we saw before,  $e^{i\Phi (t)} = \det {\cal O}_v (t)$. To
express $e^{i\Phi (t)}$ in terms of $K(t)$ we isolate ${\cal O}_v^\dagger (t)$
\begin{equation}
v^\dagger (t_1 ) v(t){\cal O}_v^\dagger (t )= v^\dagger (t_1 ) K(t) v(t_1 ), 
\end{equation}
which leads, after adding and subtracting $1=v^\dagger (t_1 ) v(t_1 )$
to
\begin{equation}
e^{-i\Phi (t) } = {{\det [1-P(t_1 ) +P(t_1 ) K(t)]}\over {\det v^\dagger (t_1 )
v(t) }}.
\end{equation}
Since $v(t)$ is single valued, $v(t_1 )=
v(t)$ for a {\it closed path} $C$, making the denominator 
unity and producing equation (6.6) of \cite{Lnonabel}.

Similarly, observing that $M_R (t) {\cal O}_v^\dagger (t) 
M_R^\dagger (t_1 )= v^{\prime\dagger} P(t) K(t) P(t_1 ) v^\prime =
v^{\prime\dagger} D_o (t) K(t) D_o^\dagger (t_1 ) v^\prime$,
we get
\begin{equation}
\det M_R (t) \left ( \det M_R (t_1 ) \right )^* =
\det [ 1-P^\prime + P^\prime D_o (t ) K(t) D_o^\dagger (t_1 ) ]
\det {\cal O}_v (t),
\end{equation}
which is equation (31) in \cite{Lpisa}.

\subsection{Connection to Continuum (simplified)}

As long as we 
focus only on one open path there is no fundamental distinction between
$|v (t ) \rangle$ and $|v_\a (t ) \rangle$. But, when we recall the
environment in which the path is embedded, it is clear that the
states $|v (t ) \rangle$ come from a single valued function on
the entire space of gauge fields. The state $|v_\a (t ) \rangle$
typically does not return to its initial value when
transported round a {\it closed} path $C$ in gauge field space. The
extra phase it acquires is Berry's phase $e^{i\Phi (C)}$. 
If $e^{i\Phi (C)}=1$ for any closed curve $C$ in gauge field space, 
a natural global phase choice would be to parallel transport by Berry's
connection from one fixed point in field space. 
Since the exact structure of the reflection $\epsilon$ is not
important for the continuum limit, but does determine $\Phi (C)$,
it is natural to search for a deformation of $\epsilon$ (``improvement'')
so that $e^{i\Phi (C)}=1$ for all closed curves $C$. It can be shown
that {\it it is impossible} to find such a deformation if continuum
perturbative anomalies {\it do not cancel} \cite{Ngeom}. It is
conjectured that if anomalies cancel and the above obstruction is removed
a deformed $\epsilon$ with $e^{i\Phi (C)}\equiv 1$ exists and would
permit a gauge invariant phase choice by Berry's law of
parallel transport \cite{Ngeom} (for a recent review see \cite{Nchiral}). 

The above conjectures are compatible with continuum: There,
nonabelian anomalies cancel iff one can make the conserved and covariant
currents identical, in other words when $\Delta J$ can be made to 
vanish. In the overlap framework the role of $\Delta J$ is played
by Berry's connection and this motivates the conjectures.
When anomalies do not cancel, the inevitability of nontrivial Berry
phases was established in \cite{Ngeom} by exhibiting a class
of backgrounds on a torus {\it in gauge orbit space} over which Berry's
curvature generated a two form which was well defined but integrated
to a non-zero integer. The result was in agreement with known continuum
formulae for $\Delta J$ \cite{Ball}. In the continuum, once
$\Delta J$ is known (including normalization), the anomalies themselves
(both consistent and covariant) are completely determined. 
The idea to add two continuous directions (in the case of \cite{Ngeom}
these were the two torus coordinates) to produce integrands that
integrate to integers (which can be non-zero if anomalies do not cancel,
as was found in \cite{Lnonabel}) 
is at the core of section 9 in \cite{Lnonabel}.

\section{Conclusions}
All recent algebraic relations are identities directly
obtainable from the basic overlap formulae. 
The one new idea is an approach to look for a gauge variant
local functional to be added to the phase of an assumed given smooth
section of the initial $U(1)$ overlap bundle so that gauge invariance
holds when anomalies cancel. The new approach is an
alternative to the older proposal of $\cite{Ngeom}$ and mathematically 
differs in replacing the geometrical framework 
of \cite{Ngeom} by one based on analysis. 
Although both approaches include some fine tuning, in
$\cite{Ngeom}$ the fine tuning is at the level 
of gauge covariant operators only. Space limitations 
prevent further discussion.

\section{Acknowledgments}
I wish to express my thanks to the organizers for inviting me to
participate in this workshop, their 
tremendous hospitality, and the pleasant atmosphere they helped create.
My research is supported in part by the DOE under grant \# DE-FG05-96ER40559.

\end{document}